\documentclass[usenatbib]{mn2e}
\usepackage{graphicx}
\usepackage{amsmath}

\def\apj{\rm ApJ}
\def\apjl{\rm ApJL}
\def\apjs{\rm ApJS}
\def\aj{\rm AJ}
\def\aaps{\rm A\&AS}
\def\mnras{\rm MNRAS}
\def\nat{\rm Nature}

\def\pasp{\rm PASP}
\def\pasa{\rm PASA}
\def\aap{\rm AAP}
\def\araa{\rm ARA\&A}

\def\physrep{\rm Physics Reports}
\def\procspie{\rm Proc. SPIE}

\def\gax{\mathrel{\raise.3ex\hbox{$>$}\mkern-14mu\lower0.6ex\hbox{$\sim$}}}
\def\lax{\mathrel{\raise.3ex\hbox{$<$}\mkern-14mu\lower0.6ex\hbox{$\sim$}}}
\def\gtorder{\mathrel{\raise.3ex\hbox{$>$}\mkern-14mu
             \lower0.6ex\hbox{$\sim$}}}
\def\ltorder{\mathrel{\raise.3ex\hbox{$<$}\mkern-14mu
             \lower0.6ex\hbox{$\sim$}}}

\begin{document}

\title [Quiescent SNe progenitors]{The quiescent progenitors of four Type II-P/L supernovae}

\author[S. A. Johnson, C. S. Kochanek, and S. M. Adams]{ 
    Samson~A. Johnson$^{1}$\thanks{Email: johnson.7080@osu.edu},
    C.~S. Kochanek$^{1,2}$,
    S.~M. Adams$^3$
    \\
  $^{1}$ Department of Astronomy, The Ohio State University, 140 West 18th Avenue, Columbus OH 43210 \\
  $^{2}$ Center for Cosmology and AstroParticle Physics, The Ohio State University, 191 W. Woodruff Avenue, Columbus OH 43210 \\
  $^{3}$ Cahill Center for Astrophysics, California Institute of Technology, Pasadena, CA 91125\\
   }

\maketitle

\begin{abstract}
We present Large Binocular Telescope difference imaging data for the final years of four Type~II-P/L supernovae progenitors.
For all four, we find no significant evidence for stochastic or steady variability in the \textit{U, B, V,} or \textit{R}-bands. 
Our limits constrain variability to no more than roughly 5-10\% of the expected \textit{R}-band luminosities of the progenitors.
These limits are comparable to the observed variability of red supergiants in the Magellanic Clouds. 
Based on these four events, the probability of a Type~II-P/L progenitor having an extended outburst after Oxygen ignition is $<37\%$ at 90\% confidence. 
Our observations cannot exclude  short outbursts in which the progenitor returns to within $\sim10\%$ of its quiescent flux on the time scale of months with no dust formation.

\end{abstract}


\section{Introduction}
\label{sec:introduction}

Stars with masses $\gax8\text{M}_\odot$ end their lives when their iron core becomes unstable and collapses.
In most cases, this results in a supernova (SN) explosion, but it is likely that 10-30\% form a black hole without a SN \citep{kochanek2008, gerke2015, adams2017}.
It has generally been assumed that the progenitors of SNe show no signs of their imminent demise, but this view has come under question from observations over the last decade.

Optical outbursts have been observed from a small number of progenitors \citep[e.g.,][]{pastorello2007,fraser2013,mauerhan2013,ofek2016}.
Others, such as Type IIn SNe, are inferred to have increased mass-loss rates near the ends of their lives based on the interactions of their ejecta with a dense circumstellar medium (CSM) \citep[e.g.,][]{galyam2012,kiewe2012, ofek2014,margutti2017}.
There are also reports of significant pre-SN activity in a growing number of otherwise normal Type II-P SNe, including SN~2007od \citep{andrews2010}, SN~2009kf \citep{botticella2010}, SN~2013fs \citep{yaron2017}, and SN~2016bkv \citep{hosseinzadeh2018}.
That these phenomena occur in the last few years before explosion requires a causal relation between the outbursts and the final phases of nuclear burning \citep[see the discussion in ][]{kochanek2011}.
Explanations for these events include instabilities in late stage nuclear burning \citep{smith2014,woosley2015} or the coupling of gravity waves in the core to the surface layers of the star \citep{quataert2012,shiode2014,fuller2017a,fuller2017b}. 

We know very little about the pre-SN variability of RSGs in terms of either luminosity or mass loss. 
Locally observed RSGs show  $\sim$10\% luminosity variations on short timescales (see Section \ref{sec:asassn} below), but can have much larger luminosity changes on longer timescales \citep[e.g.,][]{kiss2006}.
SN surveys can detect extreme outbursts of $\Delta L\gax10^{6-7}\text{L}_\odot$, which corresponds to fractional changes of $\gax1000\%$ \citep[e.g.,][]{ofek2014}.
RSGs typically lose mass at rates of $\dot{M}\lax10^{-6}$M$_\odot$/yr \citep[e.g.,][]{lamers1999}, while Type~IIn SN properties require mass loss rates of $\dot{M}\gax10^{-2}$M$_\odot$/yr leading up to the explosion \citep{kiewe2012}.
This leaves two or three orders of magnitude of possible changes in luminosity or mass loss that are unexplored yet would certainly qualify as significant pre-SN outbursts if they occurred. 
Furthermore, the extreme events which are found are rare and so tend to be distant.
This leads to difficulties in establishing a mapping between outbursts and progenitor properties such as luminosity and mass. 

\begin{figure*}
\includegraphics[width=\linewidth]{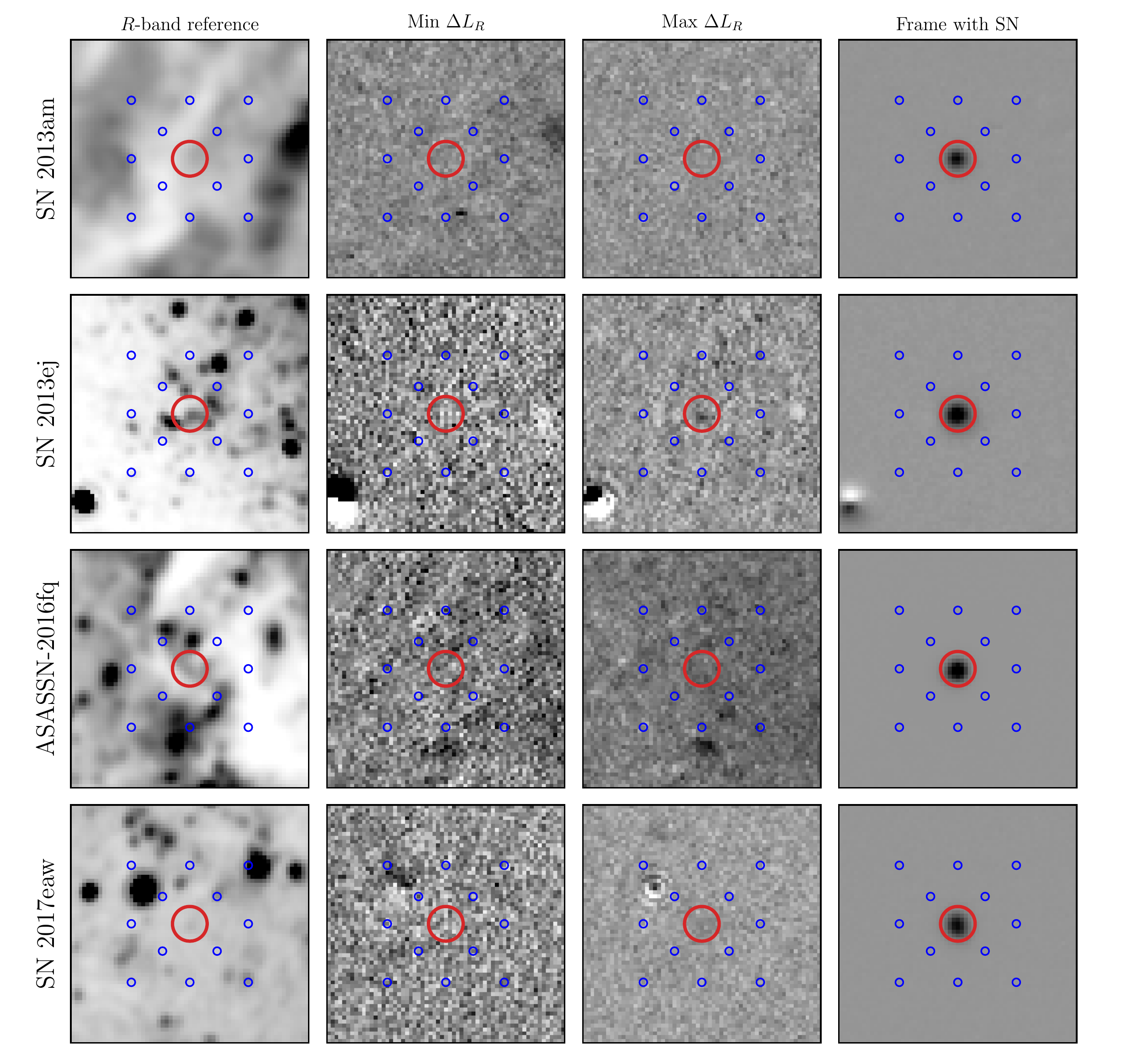}
\caption{
\textit{R}-band reference and difference images for the four progenitors. 
The first column contains the reference images, all on different scales.
The second and third columns, which are on the same scale for all four objects, show the largest luminosity excess and deficit in $\Delta\nu L_\nu$ of the higher quality data relative to the reference image.
The fourth column is a difference image containing the SN. 
The larger, red circles are $1\farcs0$ in radius, and the smaller, blue circles indicate the positions of the comparison sample pixels.
In all panels, darker colors indicate brighter sources.
}
\label{fig:four_diff}
\end{figure*}

Understanding the prevalence of pre--SN activity requires data sensitive to lower levels of progenitor variability. 
At present, this is only possible in cases where multiple \textit{Hubble Space Telescope} (\textit{HST}) observations happen to exist \citep{elias-rosa2009,fraser2014,maund2015} or for SNe that occur in the galaxies used in the Large Binocular Telescope (LBT) search for failed SNe \citep{ szczygiel2012,kochanek2017a, johnson2017}.
To date, all these examples have been essentially quiescent. 

Here, we present upper limits on the variability of one Type~II-P/L and three Type~II-P SNe progenitors in the LBT survey: SN 2013am in NGC 3623, SN 2013ej in NGC 628, ASASSN-2016fq (SN 2016cok) in NGC 3627, and SN 2017eaw in NGC 6946.
The progenitors of Type~II SNe are known to be red supergiants (RSGs) from both the presence of Hydrogen in their explosion spectra and the direct identification of their progenitors \citep[see the review by][]{smartt2015}.
Our observations and methods are summarized in Section \ref{sec:observations}.
In Section \ref{sec:progs}, we discuss the variability of each progenitor, and Section \ref{sec:asassn} examines the variability of RSGs in the Large and Small Magellanic Clouds for comparison.
We discuss our results in Section \ref{sec:discussion}.

\begin{figure*}
\includegraphics[width=\linewidth]{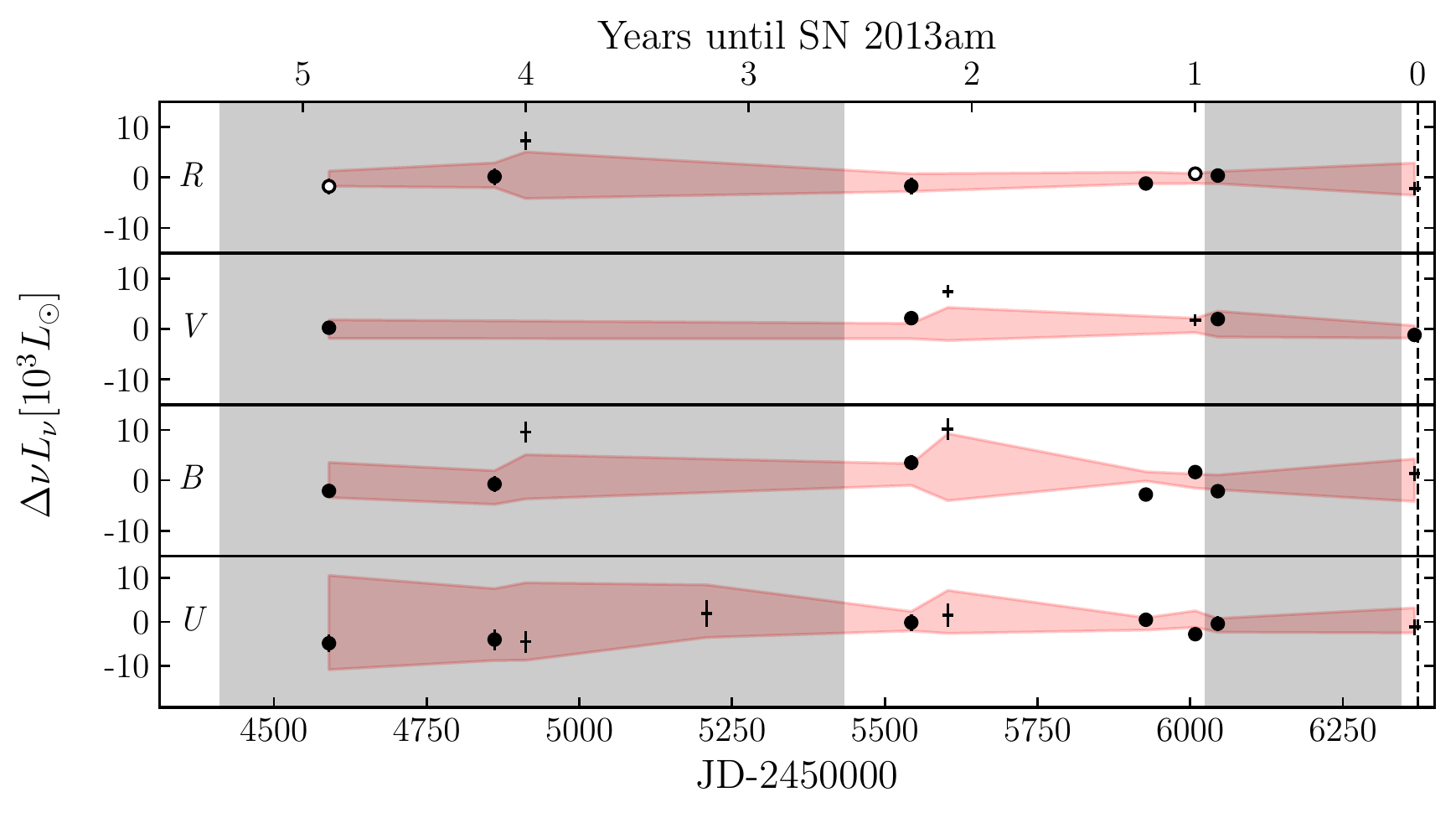}
\caption{
The \textit{UBVR} light curves of the progenitor of SN 2013am. 
The circles (crosses) represent the high (low) quality  data points. 
The filled white circles indicate the epochs displayed in Figure \ref{fig:four_diff}.
The red region is the RMS spread about the mean of the comparison light curve sample for each epoch.
The grey ranges denote the timespan in which a 12-15 M$_\odot$ star with Solar metallicity would begin core Oxygen burning (5.4-2.6 yr) and begin core Silicon burning (0.95-0.07 yr) \citep[][]{sukhbold2014,sukhbold2017}.
Be aware that the $\Delta\nu L_\nu$ scales of Figures \ref{fig:13am_pre}-\ref{fig:17eaw_pre} differ.
}
\label{fig:13am_pre}
\end{figure*}

\section{Observations}
\label{sec:observations}
The LBT survey for failed SNe began in 2008 and  can be used to study the progenitor of any subsequent successful SN in the 27 sample galaxies.
We use the Large Binocular Cameras \citep{giallongo2008} on the LBT \citep{hill2006} to image the galaxies in the \textit{U, B, V,} and \textit{R}-bands.
The data reduction and image subtraction processes using \texttt{ISIS} \citep{alard1999,alard2000} and \texttt{DAOPHOT} \citep{stetson1987} are described by \citet{gerke2015} and \citet{adams2017}.
We build our reference frames similarly to \citet{johnson2017}, using only our best pre--SN images.
This isolates the pre-SN variability of the progenitor in the difference images.
The \textit{UBVR} reference images for each host galaxy are interpolated and aligned to a common astrometric solution.

We are able to be more selective in choosing the images used to construct the reference frames for more recent SNe.
Our initial criteria for images to be used in constructing the reference frames are that they belong to the best 10 percent of images in both seeing and background flux across all 27 galaxies for each filter. 
For seeing, these values are $1\farcs15$, $1\farcs13$, $1\farcs14$, and $1\farcs10$ for \textit{U, B, V}, and \textit{R}, respectively.
For SNe that occurred earlier in the survey data, we have fewer images we can use in building the reference and thus had to relax our limits.
In particular, we needed to increase the seeing limits for images used in the reference frame of the galaxy hosting SN~2013am to $1\farcs47$, $1\farcs36$, $1\farcs37$, and $1\farcs30$. 
The limits on the background sky fluxes were the same for all four progenitors.
The reference frames are shown as the first column in Figure \ref{fig:four_diff}.
We calibrate our images using SDSS photometry \citep{ahn2012}, converted from \textit{ugriz} to \textit{UBVR} following \citet{jordi2006}.
For all four SNe, we are able to accurately determine the position of the progenitor using post-explosion images that include the fading SN.
Frames containing the SNe are shown in the fourth column of Figure \ref{fig:four_diff}.

Following \citet{johnson2017}, we place a grid of 12 trial points around each SN. 
The light curves of these comparison samples are used to examine the systematic errors in our variability estimates of the progenitors. 
The outer points of the grid are placed 15 pixels apart, which is $\sim3\farcs5$ at the LBC's plate scale of $0\farcs2255\text{ pixel}^{-1}$.
The inner points  have a spacing of 7 pixels.
Figure \ref{fig:four_diff} includes the locations of these grid points as blue circles.
We extract light curves centered on the  progenitors and the grids at each epoch using the standard PSF-weighted estimates measured by \texttt{ISIS}.
We use a PSF created from the reference frame using \texttt{DAOPHOT} when the \texttt{ISIS} generated PSF was corrupted by saturated stars.

We present data for all epochs with a seeing FWHM $<2\farcs0$, but we flag lower quality epochs defined by a seeing FWHM $>1\farcs5$ or an \texttt{ISIS} flux scaling factor $<0.8$.
A low flux scaling factor indicates that the image was obtained at a higher than average airmass or through cirrus clouds. 
The second and third columns of Figure \ref{fig:four_diff} show the epochs of higher quality data where the progenitor had the largest luminosity excess and deficit in $\Delta L_R$ compared to the reference image.
We list the number of higher quality points in each band as $N_\text{g}$ in Table \ref{tbl:var}.

\begin{figure*}
\includegraphics[width=\linewidth]{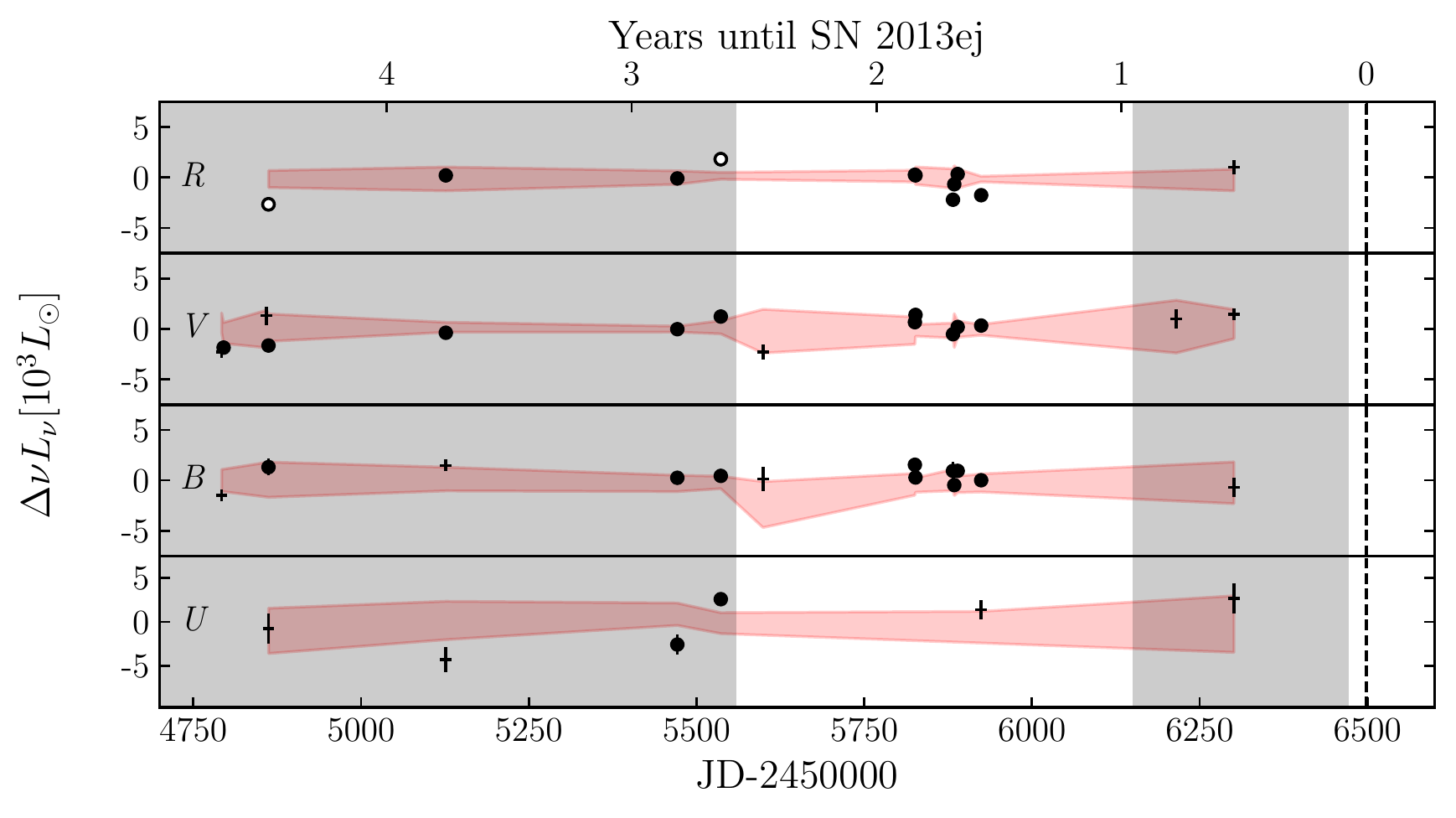}
\caption{
The \textit{UBVR} light curves of the progenitor of SN 2013ej.
The format is the same as Figure \ref{fig:13am_pre}.
Be aware that the $\Delta\nu L_\nu$ scales of Figures \ref{fig:13am_pre}-\ref{fig:17eaw_pre} differ.
}
\label{fig:13ej_pre}
\end{figure*}

\section{Individual Progenitors}\label{sec:progs}

In this section, we summarize the known properties of the progenitor to each SN and how we characterize their variability.
We use the same distances to the host galaxies as \citet{gerke2015}, the Galactic extinction from \citet{schlafly2011}, and include any estimates of the local extinction.
We assume a foreground reddening law of $R_V=3.1$ for all cases. 

To examine the variability, we utilize both the high and low quality data.
First, we characterize the ``stochastic'' variability of these SN progenitors using the root-mean-square (RMS$_p$) and peak-to-peak (PtoP$_p$) luminosity changes ($\Delta\nu L_\nu$) for each band.
The RMS$_p$ calculations include all data, while the PtoP$_p$ luminosity change estimates only include higher quality data.
The PtoP$_p$ value will not be a good measure of the variability for progenitors with few higher quality points (e.g., \textit{U}-band for SN~2013ej).
We calculate the same quantities for the comparison sample (RMS$_i$, PtoP$_i$) and report their means ($\langle$RMS$_i\rangle$, $\langle$PtoP$_i\rangle$) and dispersions in Table \ref{tbl:var}.

Both the RMS$_p$ and the PtoP$_p$ values are a combination of intrinsic variability and noise. 
We can estimate the intrinsic RMS variability of the progenitor by subtracting either the \texttt{ISIS} noise estimate $\langle \sigma^2\rangle^{1/2}$ or the average $\langle$RMS$_i\rangle$ of the comparison sample in quadrature, where the former is more conservative while the latter is likely more realistic.
These noise corrected estimates of the intrinsic variability are always non-zero if $\langle \sigma^2\rangle^{1/2}$ is used as the noise estimate, while using $\langle$RMS$_i\rangle$ as the noise estimate can drive the estimate of the intrinsic variability to be zero.
The PtoP$_p$ statistic also has some expected level of noise. 
We used Monte Carlo calculations to determine the mean contribution of Gaussian noise fluctuations to the PtoP statistics as a function of the number of data points used in the estimate. 
For example, with 4, 8, or 16 points, the mean PtoP values found for a Gaussian of width $\sigma$ are 2.1$\sigma$, 2.8$\sigma$, and 3.5$\sigma$, respectively.
We correct the PtoP$_p$ value by again subtracting this expected noise in quadrature. 

While Table \ref{tbl:var} reports variability statistics for all four bands, we will primarily discuss the \textit{V} and \textit{R}-band results.
RSGs have little blue flux and the bluer bands, especially \textit{U}-band, have more systematic problems. 
We will use the RMS$_p$ corrected by $\langle \sigma^2\rangle^{1/2}$ for our standard estimates (`Var' in Table \ref{tbl:var}) and average the \textit{V} and \textit{R}-band values for our formal limits, $\overline{\text{Var}}$. 
We also report the PtoP$_p$ corrected by $\langle \sigma^2\rangle^{1/2}$ scaled for the number of epochs used in its calculation as PtoP$_C$ in Table \ref{tbl:var}.
This correction drives some values of PtoP$_C$ to zero.

\begin{figure*}
\includegraphics[width=\linewidth]{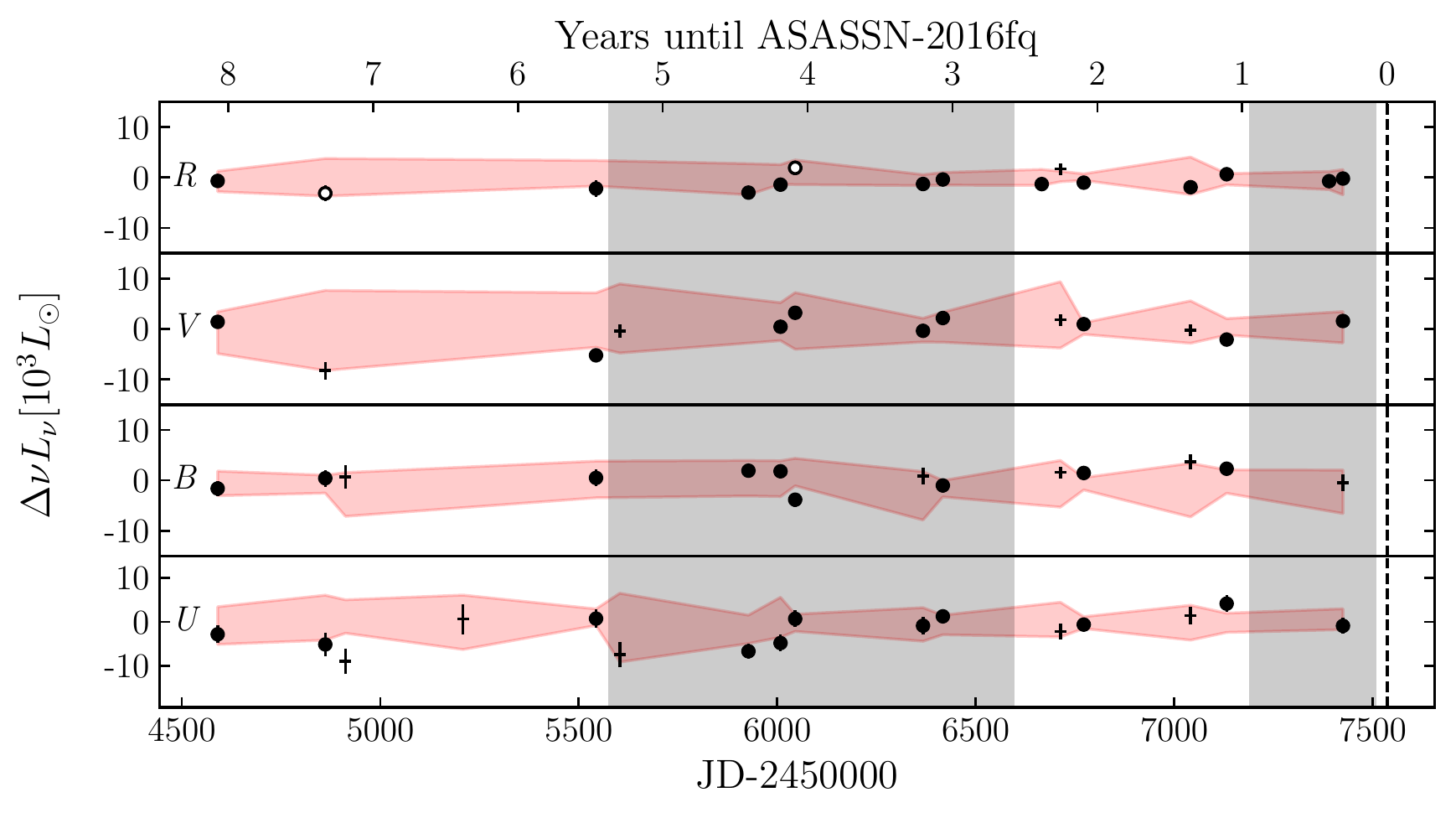}
\caption{
The \textit{UBVR} light curves of the progenitor of ASASSN-2016fq.
The format is the same as Figure \ref{fig:13am_pre}.
Be aware that the $\Delta\nu L_\nu$ scales of Figures \ref{fig:13am_pre}-\ref{fig:17eaw_pre} differ.
}
\label{fig:16cok_pre}
\end{figure*}

We investigate any long-term trends in the luminosity of the progenitors by fitting the changes in the band-luminosities with a simple line, $L(t) = A_pt+B_p$.
We also perform the same fit to the comparison sample.
For the progenitors, we report the errors in $A_p$ and for the comparison sample we report the average absolute value $\langle|A_i|\rangle$ and the standard deviations of the $|A_i|$ values about $\langle|A_i|\rangle$. 
We report the weighted average of the \textit{V} and \textit{R}-band $A_p$ values as the trend in luminosity, $\overline{A}$, although any trends in the progenitors' luminosities are consistent with the comparison sample and (typically) zero. 

\begin{figure*}
\includegraphics[width=\linewidth]{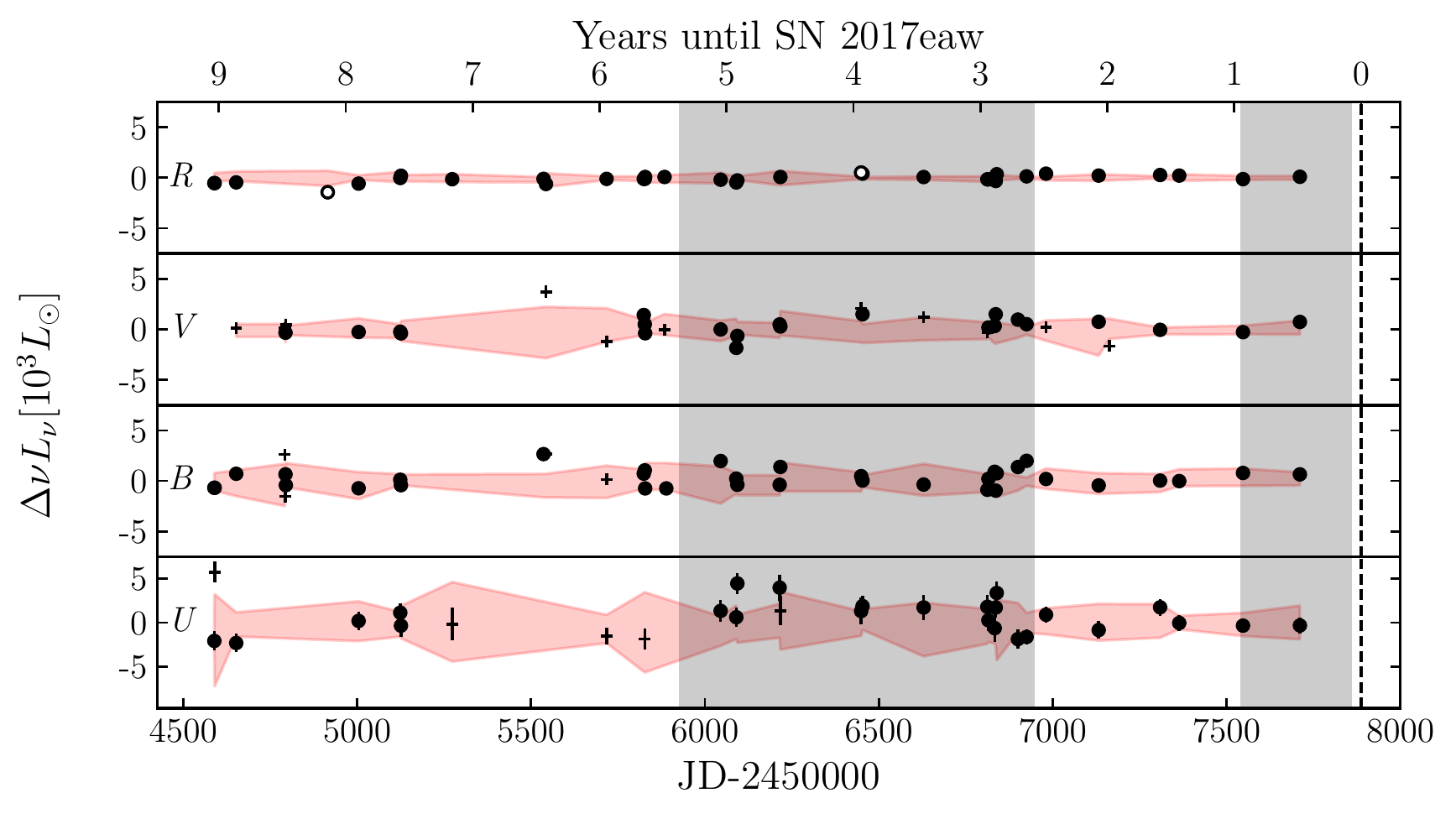}
\caption{
The \textit{UBVR} light curves of the progenitor of SN 2017eaw.
The format is the same as Figure \ref{fig:13am_pre}.
Be aware that the $\Delta\nu L_\nu$ scales of Figures \ref{fig:13am_pre}-\ref{fig:17eaw_pre} differ.
}
\label{fig:17eaw_pre}
\end{figure*}

Our differenced light curves are shown in Figures \ref{fig:13am_pre}-\ref{fig:17eaw_pre}. 
We show the changes of $\Delta\nu L\nu$ in the band luminosity $\nu L\nu$ relative to the difference image given the assumed distance and contribution of both Galactic and local extinction. 
In each figure, the luminosity scales are the same for all filters, but the scales differ from figure to figure. 
The black circles indicate the data that meet our `higher quality' criteria and the crosses indicate those that do not.
The filled white circles indicate the epochs shown in Figure \ref{fig:four_diff}.
The horizontal red shaded region is the RMS of the comparison sample about their mean for that particular epoch (this is different from $\langle$RMS$_i\rangle$).
The vertical gray bars denote the range in time before explosion in which a Solar metallicity progenitor between 12-15 M$_\odot$ begins core Oxygen burning (5.4-2.6 yr) and begins core Silicon burning (0.95-0.07 yr) \citep{sukhbold2014,sukhbold2017}.

\subsection{SN~2013am}\label{ssec:sn2013am}

The Type II-P SN 2013am was discovered independently by \citet{nakano2013} and \citet{yaron2013} on 2013-03-21.
\citet{yaron2013} and \citet{benetti2013} both classified it as a young Type II SN.
The Galactic extinction is $E(B-V)=0.02$ and \citet{zhang2014} estimate an extinction of $E(B-V)_\text{host}=0.55\pm0.19$ mag local to the SN.  
SN 2013am was observed extensively by \citet{zhang2014} in the optical and ultraviolet, who identified it as a low-velocity SN.
Based on the strength of the Calcium II features in the explosion spectra, \citet{zhang2014} suggest that SN~2013am had a relatively low mass progenitor. 
The host of this SN was NGC 3623, which has a distance of 10.62 Mpc \citep{kanbur2003}.

We had observed NGC 3623 for $\sim$5 yrs with about 8 epochs for each filter and our difference photometry for SN 2013am is shown in Figure \ref{fig:13am_pre}.
This was our least sampled progenitor, and has the some of the largest variability limits.
There are outliers in the $V$ and $B$-bands about 2 years before the SN, but these deviations are seen in the comparison sample as well.
The stochastic variability is consistent with the comparison sample, for which we place a limit of $\overline{\text{Var}}\lax2500$ L$_\odot$ from the average of the \textit{V} and \textit{R}-bands.
Any brightening or dimming trends also appear to be consistent in magnitude with those of the comparison sample.
The \textit{V} and \textit{R}-band slope estimates are both negative, but the average of the two, $\overline{A}=(-420\pm590)\text{L}_\odot/\text{yr}$, is still consistent with zero.
The \textit{U} and \textit{B}-band slopes are also consistent with zero.
 
The final pre-SN image was taken $\sim5$ days prior to the discovery of SN~2013am.
Although the photometry of this epoch was low quality, no significant variability is observed.
SN~2013am was likely quiescent until the day it died. 

\subsection{SN~2013ej}\label{ssec:sn2013ej}

\citet{kim2013} discovered  SN 2013ej in NGC 628 on 2013-07-27 and it was classified as a young Type II by \citet{valenti2013}.
\citet{yuan2016} argue for a classification of Type II-L rather than II-P. 
\citet{fraser2014} estimated a progenitor mass of 8 to 15.5 M$_\odot$ based on archival \textit{HST} images, consistent with an estimate of 12-15 M$_\odot$ based on models of nebular phase spectra by \citet{yuan2016}.
\citet{fraser2014} also present evidence of a blue source that is spatially coincident with the progenitor that dominates shorter wavelength filters. 
The progenitor should still dominate the \textit{R}-band flux in the pre-SN images.
Although their cadence is sparse, \citet{fraser2014} detect no variability above the 0.05 mag level between late 2003 and early 2005 in \textit{F814W}.
Both \citet{morozova2017} and \citet{das2017} have modeled the light curve of SN~2013ej with the inclusion of a dense CSM surrounding the progenitor. 
We adopt a distance of $9.1$ Mpc, and a Galactic extinction of $E(B-V)=0.0597$.
The equivalent width of the Na I D absorption feature of the SN suggests negligible extinction local to the progenitor \citep{bose2015}.

The host galaxy was observed with LBT $\sim$10 times over the 5 years before the SN.
The light curves for the progenitor are shown in Figure \ref{fig:13ej_pre}.
The RMS variability of the progenitor is consistent with the comparison sample, leading us to conclude that there is no significant variability in the \textit{V} and \textit{R}-bands with $\overline{\text{Var}}\lax1200 L_\odot$. 
While the slope fit in \textit{R}-band was similar to its comparison sample and zero, \textit{V} band was not.
This drives our reported linear trend in the luminosity of the progenitor to $\overline{A}=(620\pm200)\text{L}_\odot/\text{yr}$.
However, the lack of a significant slope at the adjacent \textit{B} and \textit{R}-bands suggests that the significance of the \textit{V}-band slope is overestimated. 
In terms of mass loss, \citet{morozova2017} present two extremes for the formation of the CSM of SN~2013ej: a mass loss rate of 0.2 M$_\odot$/yr (2.0 M$_\odot$/yr) for 3.12 yr (3.80 months). 
Our observations exclude the more modest wind, but the late-time, short-duration of the more extreme alternative would have not been detected in the gap between our final data point and the SN. 
If we consider the results of \citet{fraser2014} alongside our own, SN 2013ej likely had no dramatic outbursts in the last decade before its death. 

\subsection{ASASSN-2016fq}\label{ssec:asassn16}
ASASSN-2016fq (SN 2016cok) was discovered in NGC 3627 by \citet{bock2016} as part of the All-Sky Automated Survey for SNe \citep[ASAS-SN][]{Shappee2014,kochanek2017b}. 
It was classified as a Type II SN by \citet{zhang2016}.
We adopt a distance of 10.62 Mpc \citep{kanbur2003}, a Galactic extinction of $E(B-V)=0.029$, and we assume no extinction local to the SNe following \citet{kochanek2017a}. 
\citet{kochanek2017a} estimate that the progenitor was probably in the mass range of 8-12M$_\odot$.

\citet{kochanek2017a} reported the \textit{R}-band light curve and here we add the other three filters. 
Our difference photometry results for $\sim$15 observations taken over 8 yrs are shown in Figure \ref{fig:16cok_pre}.
Our \textit{R}-band values differ slightly from \citet{kochanek2017a} as we built a new reference image for this analysis. 
The apparent stochastic variability in \textit{V}-band is similar to that in the comparison sample, leading us to conclude that our measurement is limited by noise.
We find the average RMS of the \textit{V} and \textit{R}-band difference luminosities to be $\overline{\text{Var}}\lax1800\text{L}_\odot$ for the limit of stochastic variability.
Any long term trends are again consistent with noise, with an average slope of $\overline{A}=(250\pm180)\text{L}_\odot/\text{yr}$.

\subsection{SN~2017eaw}\label{ssec:sn2017eaw}

SN 2017eaw was discovered in NGC 6946 on 2017-05-14 by \citet{waagen2017} and was classified as a Type II independently by \citet{xiang2017} and \citet{tomasella2017}.
\citet{dyk2017} used archival \textit{HST} and \textit{Spitzer} observations to find that the progenitor spectral energy distribution is consistent with that of an RSG and no reddening beyond the Galactic contribution.
They also estimate an initial mass of 13 $M_\odot$ for the progenitor.
We adopt a host distance of 5.96 Mpc \citep{karachentsev2000} and a Galactic extinction of $E(B-V)=0.30$.

We observed the progenitor of SN 2017eaw about 35 times over almost 9 years before the SN.
The \textit{UBVR} light curves of the progenitor are shown in Figure \ref{fig:17eaw_pre}.
This is our most densely sampled light curve, and also showed the least variability.
We place a limit on the average \textit{V} and \textit{R}-band stochastic variability of $\overline{\text{Var}}\lax700\text{L}_\odot$, and the average of the slopes for these two bands is $\overline{A}=(70\pm20)\text{L}_\odot/\text{yr}$.

\section{ASAS-SN Variability of RSGs}\label{sec:asassn}

In order to characterize the typical variability of RSGs, we used the 226 spectroscopically confirmed K and M supergiants in the Large (\citealt{Bonanos2009}) and Small (\citealt{Bonanos2010}) Magellanic Clouds.  
We used the \textit{V}-band light curves of these stars from the All-Sky Automated Survey for Supernovae \citep[ASAS-SN,][]{Shappee2014,kochanek2017b}.  
The light curves span roughly three years with 100-200 epochs of data.  

\begin{table*}
\caption{SNe Variability limits.}
\begin{tabular}{l c c c c c r c c r r c }
\multicolumn{1}{c}{}
&\multicolumn{1}{c}{}
&\multicolumn{1}{c}{}
&\multicolumn{7}{c}{Variability $[10^3\text{L}_\odot]$}
&\multicolumn{2}{c}{Slope [$10^3\text{L}_\odot$/yr]}\\
\multicolumn{3}{c}{}
&\multicolumn{5}{c}{Progenitor}
&\multicolumn{2}{c}{Sample}
&\multicolumn{1}{c}{Progenitor}
&\multicolumn{1}{c}{Sample}\\
\multicolumn{1}{c}{SN}
&\multicolumn{1}{c}{Band}
&\multicolumn{1}{c}{$N_\text{g}$}
&\multicolumn{1}{c}{RMS$_p$}
&\multicolumn{1}{c}{$\langle\sigma^2\rangle^{1/2}$}
&\multicolumn{1}{c}{Var}
&\multicolumn{1}{c}{PtoP$_p$}
&\multicolumn{1}{c}{PtoP$_c$}
&\multicolumn{1}{c}{$\langle$RMS$_i\rangle$}
&\multicolumn{1}{c}{$\langle$PtoP$_i\rangle$}
&\multicolumn{1}{c}{$A_p$}
&\multicolumn{1}{c}{$\langle|A_i|\rangle$}\\
\hline
SN~2013am&\textit{R} &6 &$2.9$ & $1.5$ & $2.5$ & $2.5$& $0.0$ & $2.4\pm0.5$ & $4.3\pm1.1$ & $-0.46\pm0.71$ & $0.51\pm0.31$\\ 
&\textit{V} &4 & $2.7$ & $1.1$ & $2.4$ & $3.3$ & $2.5$ & $2.1\pm0.6$ & $4.4\pm1.2$ & $-0.33\pm1.09$ & $0.37\pm0.22$\\ 
&\textit{B} &6 & $4.6$ & $1.6$ & $4.3$ & $6.3$ & $4.9$ & $3.6\pm1.0$ & $6.0\pm2.3$ & $-0.24\pm1.07$ & $0.61\pm0.33$\\ 
&\textit{U} &6 & $2.4$ & $2.2$ & $0.9$ & $5.4$ & $0.0$ & $5.7\pm1.6$ & $12.6\pm4.3$ & $0.76\pm0.47$ & $2.21\pm0.99$\\ 

&&&&&&\\
SN~2013ej&\textit{R} &10 &$1.3$ & $0.6$ & $1.2$ & $4.5$& $4.1$ & $0.8\pm0.2$ & $2.3\pm0.8$ & $-0.03\pm0.56$ & $0.28\pm0.19$\\ 
&\textit{V} &10 & $1.3$ & $0.5$ & $1.1$ & $3.2$ & $2.8$ & $1.3\pm0.3$ & $2.5\pm0.8$ & $0.71\pm0.21$ & $0.27\pm0.13$\\ 
&\textit{B} &9 & $0.9$ & $0.7$ & $0.5$ & $2.0$ & $0.0$ & $1.4\pm0.3$ & $3.1\pm0.9$ & $0.24\pm0.25$ & $0.36\pm0.14$\\ 
&\textit{U} &2 & $2.6$ & $1.4$ & $2.3$ & $5.2$ & $4.9$ & $2.0\pm0.9$ & $2.2\pm1.5$ & $1.47\pm1.53$ & $0.40\pm0.35$\\ 

&&&&&&\\
ASASSN-2016fq&\textit{R} &14 &$1.4$ & $1.2$ & $0.8$ & $5.0$& $3.1$ & $2.2\pm0.8$ & $7.7\pm3.0$ & $0.24\pm0.21$ & $0.35\pm0.32$\\ 
&\textit{V} &9 & $3.1$ & $1.1$ & $2.8$ & $8.4$ & $7.7$ & $4.5\pm1.9$ & $9.6\pm3.4$ & $0.27\pm0.37$ & $0.72\pm0.46$\\ 
&\textit{B} &9 & $1.8$ & $1.5$ & $1.0$ & $6.1$ & $4.2$ & $3.6\pm1.3$ & $7.3\pm2.1$ & $0.34\pm0.25$ & $0.21\pm0.10$\\ 
&\textit{U} &11 & $3.6$ & $2.2$ & $2.8$ & $10.9$ & $8.4$ & $3.9\pm1.4$ & $9.7\pm3.8$ & $0.81\pm0.37$ & $0.43\pm0.33$\\ 

&&&&&&\\
SN~2017eaw&\textit{R} &34 &$0.4$ & $0.1$ & $0.3$ & $1.9$& $1.8$ & $0.3\pm0.1$ & $1.6\pm0.7$ & $0.07\pm0.02$ & $0.04\pm0.03$\\ 
&\textit{V} &22 & $1.0$ & $0.3$ & $1.0$ & $3.3$ & $3.1$ & $0.9\pm0.5$ & $3.0\pm2.0$ & $0.04\pm0.05$ & $0.06\pm0.06$\\ 
&\textit{B} &34 & $1.0$ & $0.5$ & $0.9$ & $3.6$ & $2.9$ & $1.1\pm0.4$ & $4.7\pm1.6$ & $-0.00\pm0.07$ & $0.06\pm0.04$\\ 
&\textit{U} &25 & $1.9$ & $1.2$ & $1.5$ & $6.8$ & $4.8$ & $2.5\pm0.3$ & $7.8\pm0.8$ & $-0.01\pm0.14$ & $0.18\pm0.13$\\ 

\end{tabular}
\label{tbl:var}
\end{table*}

Many of the light curves were obtained from several ASAS-SN cameras and we removed any (generally small) residual calibration differences using the methods from \citet{Kozlowski2010}.
We modeled the variability of the stars using a damped random walk (DRW) with a fixed $\tau=164$~day time scale, fitting for the variability amplitude $\sigma$ and constant offsets for the light curve from each camera.  
The DRW model simply provides a well-defined method of interpolation based on a rough fit to the variability power as a function of time scale (i.e., the structure function).  
There is no need for it to be a correct, physical description of RSG variability for this purpose -- it is simply a mathematical tool. 
As part of this process, the error estimates are re-scaled so that the model has a $\chi^2$ per degree of freedom of unity if the initial fit has a higher value.
The resulting offsets are then removed along with the 0.1\% of the data points which lay more than $3\sigma$ from the DRW model.

We compute the same RMS and PtoP statistics for this sample as we do for the progenitors, although the photometric errors are far less significant for these bright stars. 
The statistics in terms of magnitudes are only useful if the flux of the progenitor is known, while in most cases we lack such a measurement.
For these cases, it is more useful to examine the changes in luminosity, as these can be measured or constrained from the difference images without knowing the absolute luminosity of the progenitor.
We converted to luminosities using a distance modulus of $\mu=18.5$ and $E(B-V)=0.066$ for the LMC and $\mu=18.9$ and $E(B-V)=0.032$ for the SMC, which assumes only Galactic extinction from \citep{schlafly2011}.

\begin{figure}
\centering
\includegraphics[width=\columnwidth]{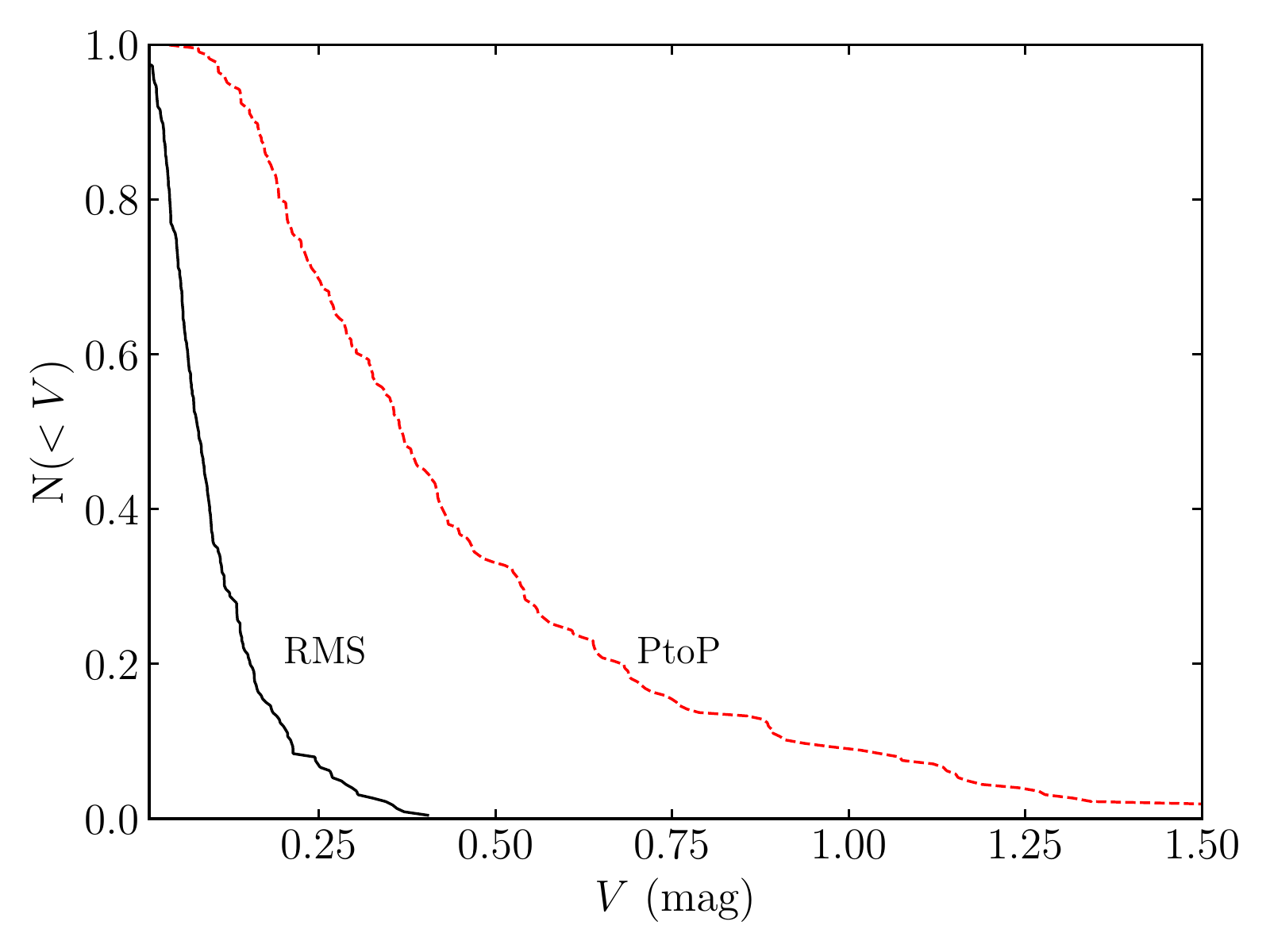}
\caption{ Integral distributions of the ASAS-SN \textit{V}-band RMS (black solid) and peak-to-peak (PtoP, red dashed) 
  magnitude variability of spectroscopically confirmed RSGs in the Magellanic Clouds from \citet{Bonanos2009,Bonanos2010}.
  }
\label{fig:mcmag}
\end{figure}

\begin{figure}
\centering
\includegraphics[width=\columnwidth]{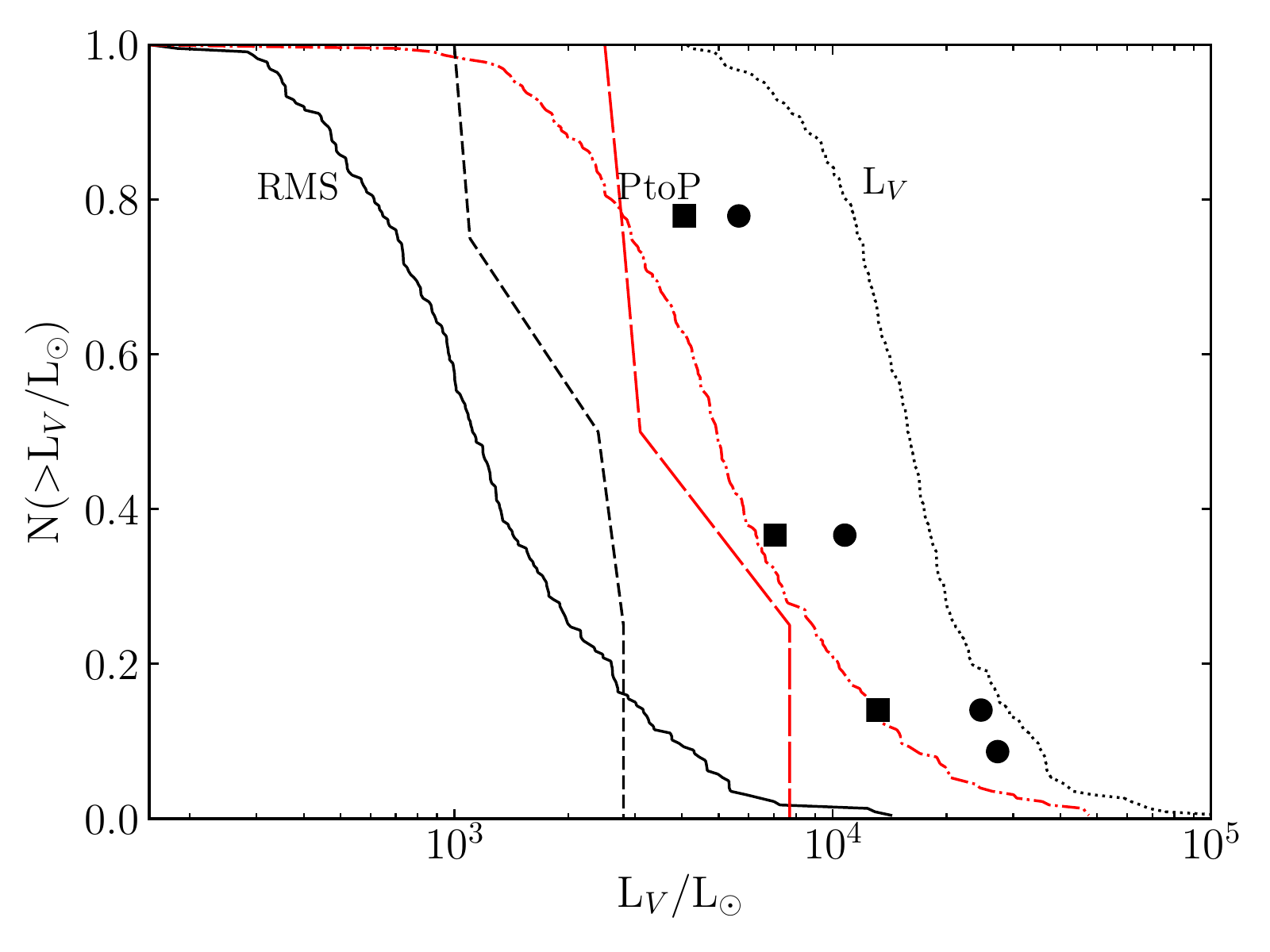}
\caption{ 
Integral distributions of the ASAS-SN \textit{V}-band RMS (black solid) and peak-to-peak (PtoP, red dash-dot) luminosity variability of spectroscopically confirmed RSGs in the Magellanic Clouds from \citet{Bonanos2009,Bonanos2010}. 
The black dotted $L_{V}$ curve shows the integral distribution of the stars in \textit{V}-band luminosity.
The black dashed (red long-dashed) line is the integral distribution of the \textit{V}-band `Var' (`PtoP$_C$') values for our four progenitors reported in Table \ref{tbl:var}.
The squares (circles) are the \textit{V}-band luminosities from the non-rotating (rotating) pre-explosion RSG models by \citet{groh2013} distributed in fraction to follow a Salpeter initial mass function from 8 to 18M$_\odot$.
}
\label{fig:mclum}
\end{figure}

Figure~\ref{fig:mcmag} shows the integral distribution of RSGs in their RMS and PtoP magnitude variability after correcting for the typical amplitude of the photometric errors.  
The median RSG has an RMS \textit{V}-band variability of $0.08$~mag and a peak-to-peak amplitude of $0.37$~mag over the three years spanned by the data.  
The integral distributions for the RMS and PtoP \textit{V}-band luminosity variability are shown in Figure~\ref{fig:mclum}.  
Here the median RMS variability is $1100L_\odot$ and the median peak-to-peak amplitude is $4900L_\odot$ as compared to a median \textit{V}-band luminosity of $1.6 \times 10^4 L_\odot$.  
The median ratio of the RMS variability to the \textit{V}-band luminosity is $\sim8\%$, while the mean is $\sim10\%$.
The peak-to-peak amplitude is well-correlated with the RMS luminosity but approximately four times larger.

This is not an ideal comparison sample for the SN progenitors we consider. 
First, it is not a well-defined, complete sample of RSGs in the LMC and SMC since it simply uses the RSGs with well-defined spectral types from \citet{Bonanos2009,Bonanos2010}.  
Second, we used \textit{V}-band because it was available from ASAS-SN instead of \textit{R}-band where the LBT data are deeper and the RSGs are more luminous.

Figure \ref{fig:mclum} shows that our variability limits on the progenitors are comparable to the observed variability of RSGs in the Magellanic Clouds.
We see no evidence for excess variability in ``pre-SN'' RSGs compared to ``normal'' RSGs.
In fact, our estimates for the contribution of noise to the apparent variability of the SN progenitors were probably too small because we made the correction using $\langle \sigma^2\rangle^{1/2}$ rather than $\langle$RMS$_i\rangle$.
So if anything, ``pre-SN'' RSGs may be \textit{less} variable at \textit{V}-band than ``normal'' RSGs. 

We should also note that the \textit{V}-band luminosities of the Cloud RSGs appear to be higher than the expectations for SN progenitors. 
We illustrate this using the \textit{V}-band luminosities of the \citet{groh2013} SN progenitor models.
We construct the integral distribution assuming a Salpeter initial mass function ($dN/dM\propto M^{-2.35}$) from 8 to 18M$_\odot$ based on the results of \citet{smartt2015}.
The model progenitor \textit{V}-band luminosity distribution is shifted to luminosities 2-3 times lower than is observed for the Cloud RSGs.

\section{Discussion}\label{sec:discussion}

\begin{figure}
\centering
\includegraphics[width=\columnwidth]{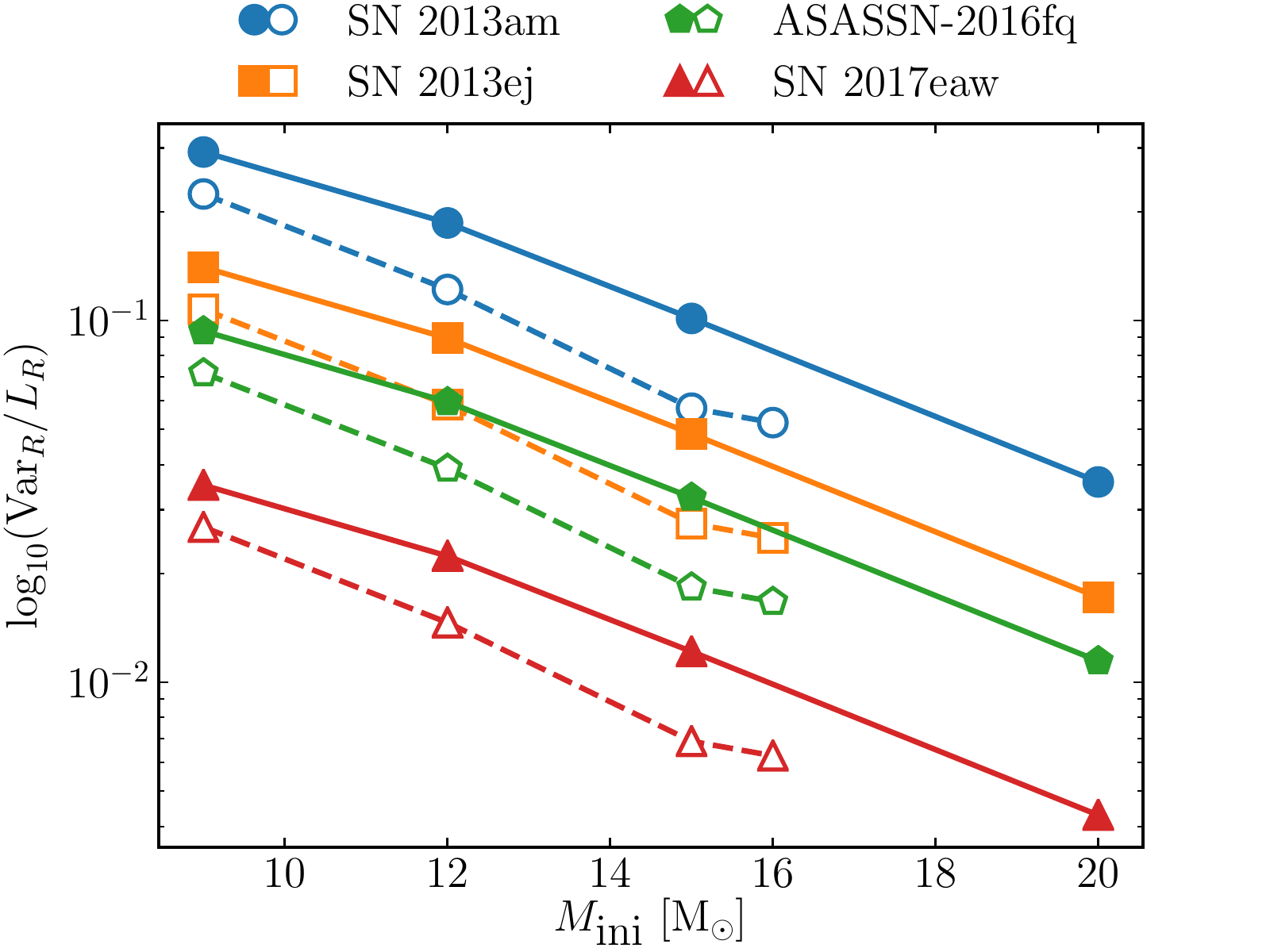}
\caption{ 
The fractional variability implied by our \textit{R}-band Var limits and the \textit{R}-band luminosity of the \citet{groh2013} progenitor models.
The filled (open) points indicate non-rotating (rotating) model values. 
}
\label{fig:frac}
\end{figure}

We have placed limits on the luminosity variability of four progenitors to Type II-P/L SNe in nearby galaxies. 
We find no convincing evidence for either stochastic variability or steady trends in the luminosity over the last several years of these progenitors lives.
As shown in Figures \ref{fig:13am_pre}-\ref{fig:17eaw_pre}, the data roughly span the period from Oxygen core ignition through death.
These SNe were fairly typical Type II's, and the limits on their variability are broadly consistent with the variability of RSGs in the Magellanic clouds.
We illustrate this further in Figure \ref{fig:frac}, where we show the implied fractional variability at \textit{R}-band as a function of progenitor mass using the \textit{R}-band luminosities of the rotating and non-rotating RSG progenitors from \citet{groh2013} since we do not have good progenitor luminosity and mass estimates for all four stars. 
For the lowest mass RSG SNe progenitor models, the variability is $\lax$20\% of the luminosity in this band, and it could be as low as $\sim$2\% for the higher mass models.
We also show the mean of the fractional variation in the LMC and SMC RSGs (Section \ref{sec:asassn}) as a black horizontal line with the standard deviation of the variations as the grey shaded region. 
Since the initial mass function favors lower mass stars, we can infer that the typical RMS variability is 5-10\% or less, which is roughly consistent with Figure \ref{fig:mcmag}.
Once the SNe of these progenitors fade, we will be able to measure/constrain their total band luminosities as described in \citet{johnson2017}.

Recent models of pre-SN outbursts of RSGs by \citet{fuller2017a} generally and by \citet{morozova2017} and \citet{das2017} specifically for SN~2013ej predict order unity luminosity changes in the final year(s) up to explosion or employ extremely elevated mass loss rates.
This is not observed for these four stars.
While sufficiently short duration outbursts of luminosity could be hidden between our observation epochs, their consequences likely could not be for two reasons.
First, while an outburst can be fast and impulsive, the return to equilibrium cannot be. 
The dynamical timescale of an RSG is long ($\sim$yr), and the thermal timescales are longer still. 
Stellar mergers are a potential analogy, where events like V838 Mon brighten quickly but fade relatively slowly \citep{munari2002}.
Second, scattering and dust formation in the ejecta would have observable effects long after the outbursts. 

For example, consider a simple dust formation model scaled by the 12M$_\odot$ non-rotating SN progenitor from \citet{groh2013} with $T_*=3500T_{*3.5}\text{K}$ and  $R_*=600R_{*0.6}\text{R}_\odot$. 
If we assume a dust formation temperature of $T_d = 1000T_{d1.0}\text{K}$ and black body temperatures, we find a dust formation radius of
\begin{equation}
\label{eqn:rad}
R_d = \frac{R_*}{2}\left(\frac{T_*}{T_d}\right)^2=3700R_{*0.6}\left(\frac{T_{*3.5}}{T_{d1.0}}\right)^2\text{R}_\odot\\
\end{equation}
\begin{equation*}
=3700R_{d3.7}\text{R}_\odot.
\end{equation*}
If we assume a wind velocity of $v_w=60v_6\text{km}/\text{s}$ scaled to the escape velocity of this progenitor, then the dust formation time scale is approximately
\begin{equation}
t_d = \frac{R_{d3.7}}{v_{w6}}\approx1.5 \text{yr}.
\end{equation}
For a typical dust opacity of $\kappa=100\kappa_{1}\text{cm}^2/\text{g}$, a mass loss rate of $\dot{M}=10^{-4}\dot{M}_{4}\text{M}_\odot/\text{yr}$, and the dust formation radius from Equation \ref{eqn:rad}, the visual optical depth scale of the dusty wind is then
\begin{equation}
\tau_{V}=\frac{\kappa \dot{M}}{4\pi R_dv_w} = 34\frac{\kappa_{1}\dot{M}_{4}}{R_{d3.7}v_{w6}}.
\end{equation}
For these cold stars, there is no difficulty forming dust \citep{kochanek2011,kochanek2014} and dense ejecta with parameters even within an order of magnitude of these scalings would have obvious photometric impacts for decades.
Even if we assume a Thomson opacity of $\kappa_\text{T}=0.4\text{cm}^2/\text{g}$, the scattering optical depth of the wind could be significant.
Essentially, dense late-time outbursts of cool stars should promptly render the star optically invisible for long periods of time independent of the transient duration.

The absence of outbursts also limits their frequency.
If $f$ is the fraction of SN progenitors having extended outbursts shortly before death, then the probability of seeing no extended outbursts in a sample of $N$ objects is $(1-f)^N$.  
For the $N=4$ Type~II SNe we consider here, the lack of extended outbursts implies that $f<0.37$ at 90\% confidence.  
By ``extended'', we mean outbursts that are sufficiently long to require no corrections for the temporal gaps between the LBT epochs or have other observable consequences, such as dust formation, long after they occur.
If we use all $N=6$ SNe in the LBT sample, adding the low-variability Type~IIb SN~2011dh \citep{szczygiel2012} and the Type~Ic SN~2012fh \citep{johnson2017}, then $f<0.28$ at 90\% confidence for SN progenitors more generally.

An alternative way of describing the limits is as a mass range. 
We again assume that Type~II SNe to come from stars with masses between $M_1=8$ and $M_2=18 M_\odot$ and assume they follow a Salpeter IMF.  
The outburst fraction could correspond to a fractional mass range over this interval if you assume that all stars of a given mass either experience outbursts or do not do so.   
If we assume that only the lowest mass stars from $M_1<M<M_o$ show outbursts, then the lack of outbursts from the 4 Type~II's implies that $M_o<9.9M_\odot$ based on the 90\% confidence limit on $f$.  
If we assume that only the highest mass stars from $M_o<M<M_2$ show outbursts, then $M_o>14.0M_\odot$. 

In some sense, we know these four SNe probably lacked the dramatic outbursts seen in SN surveys or Type~IIn SNe, as they were fairly normal Type II-P/L SNe.
Still, as discussed in the introduction, there is a huge gap between the quiescent properties of RSGs and $10^{6-7}\text{L}_\odot$ outbursts or mass loss rates $>10^{-3}\text{M}_\odot/\text{yr}$.
The key statistic for Type II's will be to associate the absence of outbursts with the luminosities and masses of the progenitors. 
While the LBT sample has yet to include a Type IIn, the number of successful SNe in the LBT survey is increasing by $\sim1$/yr.
Since Type IIn make up $\sim$9\% of Type II SNe in a volume limited sample \citep{li2011}, we will eventually be able to report on the pre-SN variability of such a progenitor for comparison to these normal Type II SNe.

\section{Acknowledgments}
We thank the referee for their comments and suggestions that helped clarify our work. 
We thank K. Stanek for discussions on RSG variability, and T. Sukhbold for providing models on SNe progenitors. 
CSK is supported by NSF grants AST-1515876 and AST-1515927. 
This work is based on observations made with the Large Binocular Telescope. 
The LBT is an international collaboration among institutions in the United States, Italy, and Germany. 
The LBT Corporation partners are: the University of Arizona on behalf of the Arizona university system; the Istituto Nazionale di Astro. 
This research has made use of the NASA/ IPAC Infrared Science Archive, which is operated by the Jet Propulsion Laboratory, California Institute of Technology, under contract with the National Aeronautics and Space Administration. 
ASAS-SN is funded in part by the Gordon and Betty Moore Foundation through grant GBMF5490 to the Ohio State University, NSF grant AST-1515927, the Mt. Cuba Astronomical Foun dation, the Center for Cosmology and AstroParticle Physics (CCAPP) at OSU, and the Chinese Academy of Sciences South America Center for Astronomy (CASSACA).

\vfill\eject

\end{document}